\documentclass[graybox, envcountchap]{svmult}

\usepackage{mathptmx}        
\usepackage{helvet}          
\usepackage{courier}         

\usepackage{makeidx}         
\usepackage{graphicx}        
\usepackage{multicol}        
\usepackage[bottom]{footmisc}

\makeindex             

\usepackage{amsmath}
\usepackage{amssymb}
\usepackage{lscape}
\usepackage{multirow}

\def\gapp{\ifmmode\stackrel{>}{_{\sim}}\else$\stackrel{<}{_{\sim}}$\fi}
\def\gsim{\lower.5ex\hbox{\gtsima}}
\def\gtsima{$\; \buildrel > \over \sim \;$}
\def\lapp{\ifmmode\stackrel{<}{_{\sim}}\else$\stackrel{<}{_{\sim}}$\fi}
\def\lsim{\lower.5ex\hbox{\ltsima}}
\def\ltsima{$\; \buildrel < \over \sim \;$}

\newcommand\apgt{\ {\raise-.5ex\hbox{$\buildrel>\over\sim$}}\ }
\newcommand\aplt{\ {\raise-.5ex\hbox{$\buildrel<\over\sim$}}\ }
\begin{document}
\pagestyle{empty}
\frontmatter


\mainmatter

%
%
%

\setcounter{chapter}{1}


\title{Blue Straggler Stars: Early Observations that Failed to Solve the Problem}
titlerunning{Early Observations} 
\author{Russel D. Cannon}
\institute{Russel D. Cannon \at Anglo Australian Observatory, PO Box 915, North Ryde, NSW 1670, Australia,\\
\email{rdc@aao.gov.au}}

%
%
\maketitle
\label{Chapter:Cannon}

\abstract*{In this chapter, I describe early ideas on blue stragglers, and various observations (some published, some not) that promised 
but failed to resolve the question of their origin. I review the data and ideas that were circulating from Allan Sandage's original discovery in 1953 of ``anomalous blue stars'' in the globular cluster M3, up until about 1992, when what seems to have been the only previous meeting devoted to Blue Straggler Stars (BSSs) was held at the Space Telescope Science Institute.
}

\section{Introduction}
\label{cansec1}
The editors asked me to summarise what must amount to prehistory for many of today's readers. My aim is to review the data and ideas that were circulating from Allan Sandage's \cite{cannsa53} original discovery of ``anomalous blue stars'' in the globular cluster M3\index{M3}, up until about 1992, when what seems to have been the only previous meeting devoted to Blue Straggler Stars (BSSs) was held at the \emph{Space Telescope Science Institute}. Fortunately, the participants at that somewhat informal ``Journal Club'' meeting persuaded the organisers that it would be worthwhile to publish the proceedings \cite{cannsa93}. I refer readers to that publication, which captures the state of BSS research twenty years ago, and in particular to the introductory paper by Mario Livio \cite{li93} and the concluding summary by Virginia Trimble \cite{tr93}. Papers by Paresce \cite{pa63} and Guhathakurta et al. \cite{gu63} reported the discovery by of new populations of BSSs in the cores of some globular clusters\index{globular cluster}, using the \emph{Hubble Space Telescope}\index{Hubble Space Telescope} in its original de-focused state, which marked a major new development in the field.

The subtitle of Livio's review \cite{li93}, ``The Failure of Occam's Razor'', essentially says it all. Several plausible and some rather implausible hypotheses had been put forward to explain BSSs, with the most popular involving binary stars transferring mass\index{mass transfer} or merging\index{merger}. It seems that many ways of making BSSs do sometimes occur in nature, although most routes are rare. Much of the confusion probably arose because the purely descriptive but unphysical term ``Blue Straggler'' came to mean different things to different people. 

A few quotations illustrate how perceptions of blue stragglers evolved over four decades. Sandage, who was involved in many of the key observations, changed his mind several times regarding their significance or even reality (Sect.~\ref{cansec2}).  Spinrad \cite{sp66} wrote, with considerable prescience:
\begin{quotation}
The few blue stragglers in M67\index{M67} are usually dismissed --- M67 would then make a perfect model for Baade's Galactic Centre Population.
\end{quotation}
Precisely how to include BSSs in modelling distant galaxies remains a key issue today. In an early review of BSSs data and explanations, Craig Wheeler \cite{wh79} wrote:
\begin{quotation}
... blue stragglers remain one of the unexplained oddities of astronomical lore. 
\end{quotation}
Several other reviews of blue stragglers appeared in the early 1990s. Trimble \cite{tr92} concluded that: 
\begin{quotation}
All in all, blue stragglers seem to be both inadequately understood and insufficiently appreciated.
\end{quotation}
Piet Hut and eight eminent co-authors \cite{hu92} focused specifically on the role of binary stars, including BSSs, in globular clusters, based on a meeting of theorists and observers at Princeton in 1991. Their abstract ends with 
\begin{quotation}
... the fascinating interplay between stellar evolution and stellar dynamics which drives globular cluster evolution.
\end{quotation}
Stryker \cite{st93} concluded that 
\begin{quotation}
Progress has certainly been made in the last 40y, but BSSs remain an intriguing challenge, \end{quotation}
while Bailyn \cite{ba95} concentrated on the critical importance of binary stars in the dynamical evolution of globular clusters. He too provided some pertinent quotes: 
\begin{quotation}... every kind of object... can be made in at least two different ways, all of which are likely to be significant...
\end{quotation} and 
\begin{quotation}
Efforts to provide simple explanations for the full range of observed phenomena appear doomed to failure. 
\end{quotation}
Evidently blue stragglers had gradually risen in status, from being either an annoying or intriguing curiosity to becoming an important factor in the evolution of star clusters and galaxies that may well have cosmological significance.

By 1992 there was no longer any doubt about the existence of blue stragglers but it was far from clear how many different formation mechanisms were involved or how many different classes of BSSs existed, while their wider significance for stellar evolution and a full understanding of stellar systems was just starting to be discussed. 
	
Given the existence of several thorough reviews published in the early 1990s, my main focus here will be on the earlier period up to about 1970. Here, I have drawn upon an unpublished chapter of my 1968 PhD thesis that dealt with the BSSs in a sample of five old open clusters\index{open cluster}. This work is described in a ``summary of a summary'' given at a Herstmonceux Conference and is sometimes referred to as ``Cannon, 1968'' although it is hardly a proper publication. This is typical of the problems faced in trying to track down even one's own publications in the pre-digital age: many papers appear in obscure journals or observatory publications that are often not accessible online. Even the origin of the name ``blue straggler'' is uncertain (\cite{tr93}: see Sect.~\ref{cansec22} below).

\section{The \emph{Classical} Blue Stragglers}
\label{cansec2}
\subsection{Globular clusters}\label{cansec21}
\begin{figure}[b]
\includegraphics[]{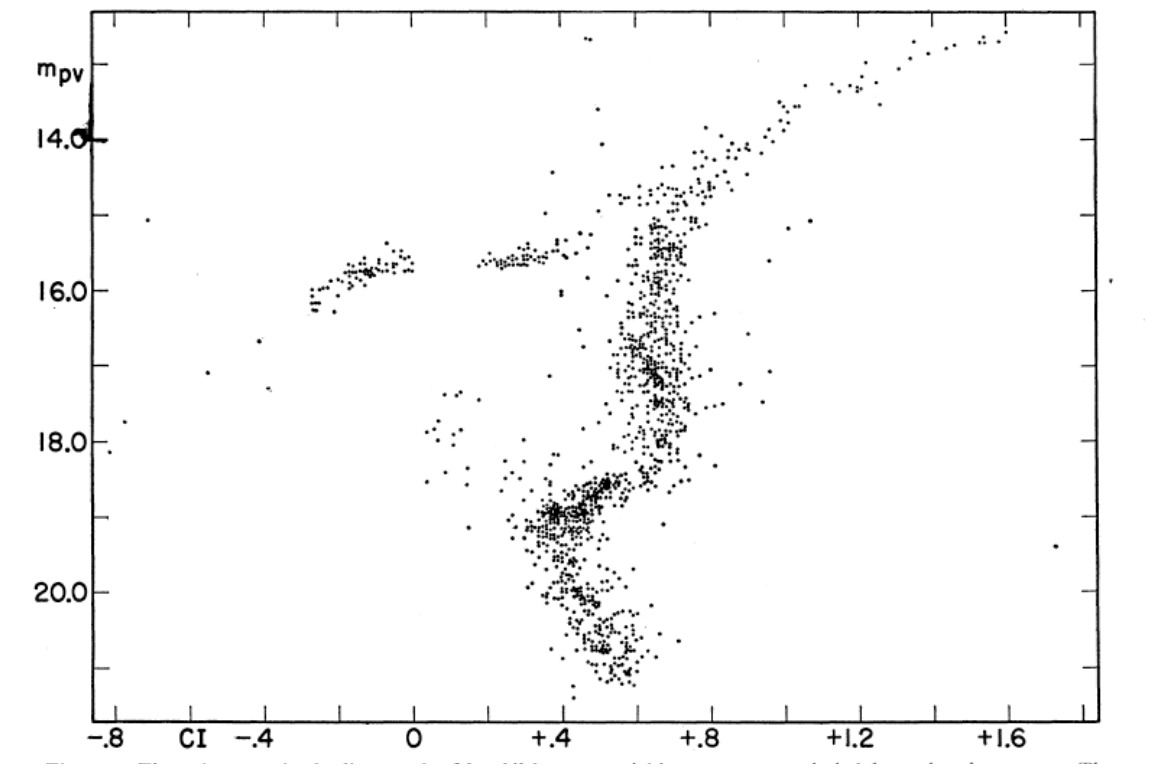}
\caption{Sandage's first CMD of the globular cluster M3 \cite{cannsa53}, showing blue stars lying above the main sequence turn-off near $m_{pv} \approx$18, CI  $\approx$ +0.2. Reprinted with permission from the AAS.}
\label{canfig1}       
\end{figure}

The discovery of BSSs came with Sandage's first Colour-Magnitude Diagram\index{colour-magnitude diagram} (CMD) of M3\index{M3} \cite{cannsa53}, part of his PhD work (Fig.~\ref{canfig1}). M3 shows all the major features that characterise the CMDs of globular clusters: it led directly to much of our understanding of stellar evolution, starting from an ``Observational Approach to Stellar Evolution'' by Sandage himself \cite{sa57}. M3 also contains a group of blue stars lying above the main sequence turn-off (MSTO), apparently younger than the bulk of the cluster stars. They did not look like a single second population of younger stars, nor did they appear related to the Blue Horizontal Branch (BHB). A minor worry was that the majority of the initial sample lay in one quadrant, a good example of the often misleading properties of small samples (\cite{eg64}: discussion between Schwarzschild and Sandage). 

For almost 20 years M3 remained the only convincing example of a globular cluster with BSSs, until Arp \& Hartwick \cite{ar71} looked at M71\index{M71}. Cannon \cite{ca68} listed eight globular clusters\index{globular cluster} with CMDs that might have revealed BSSs but only M92\index{M92} \cite{ar53} had sufficiently accurate photometry\index{photometry} for enough stars to reveal just one potential BSS, indicating a much smaller population than in M3, relative to subgiants at the same magnitudes. By 1992, the situation was very different, with lists of over 400 BSSs in more than 20 globular clusters \cite{cannsa93,fu93}.

%

\begin{figure}[b]
\includegraphics[]{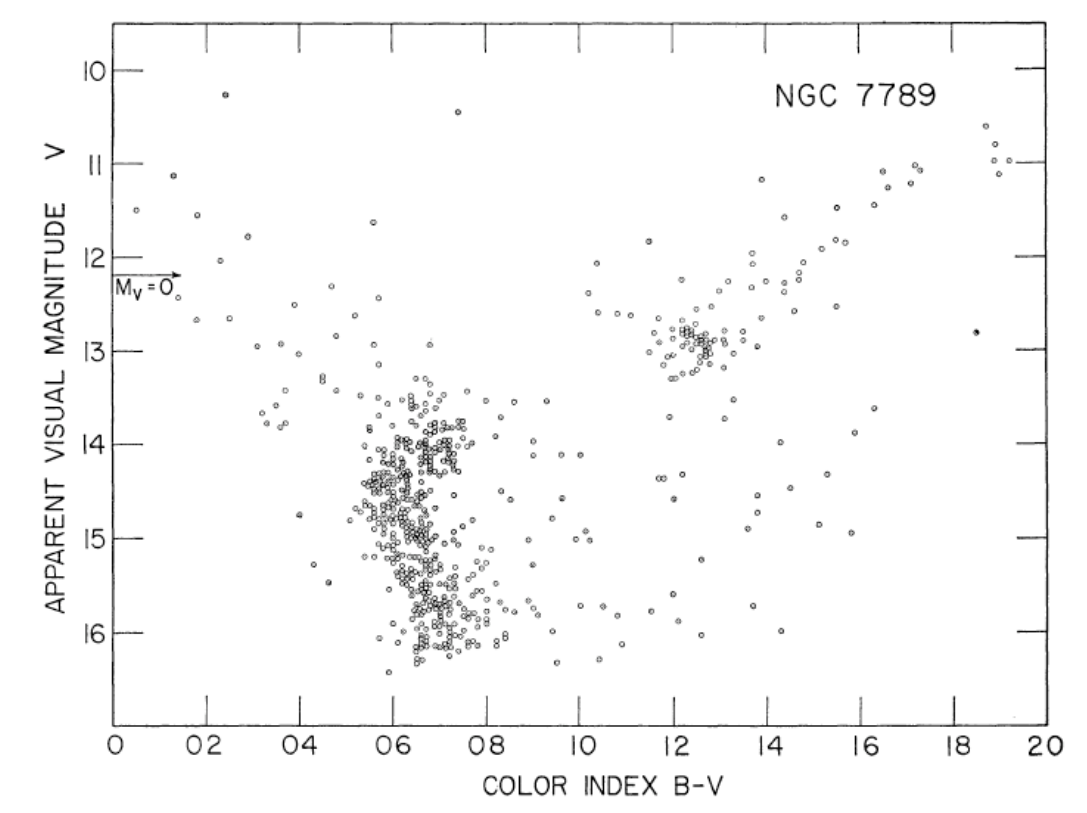}
\caption{The first CMD for a rich intermediate-age open cluster, NGC 7789 \cite{bu58}. This shows a substantial population of \textit{blue stragglers} in the upper left-hand corner, well above the main sequence turn-off. Reprinted with permission from the AAS.}
\label{canfig2}       
\end{figure}

\subsection{The Older Open Clusters}\label{cansec22}
Meanwhile, blue stars above the MSTO had been found in several intermediate-age and old open clusters, with (modern) ages in the range 0.5 to about 8 Gyrs. The earliest and still one of the best examples was NGC 7789\index{NGC 7789} \cite{bu58} with over 30 such stars, most of which seemed likely to be cluster members on statistical grounds.  As in M3, the large sample of BSSs in NGC 7789 do not lie on a well-defined sequence, as would be expected for a single much younger stellar population (familiar nowadays for example in the Carina Dwarf Spheroidal galaxy \cite{sm96}). 

As an aside, Trimble \cite{tr93} set a challenge for someone to discover a reference to the term ``blue straggler'' prior to 1965. Javier Ahumada reminded us about NGC 7789 during the Ecology of Blue Straggler Stars workshop and probably deserves the glass of wine referred to by Trimble, although George Preston who was present at both meetings was not sure who should provide the prize. A search of the NASA ADS database confirms that Burbidge \& Sandage \cite{bu58} were the first to use the name in print. My personal recollection is that ``blue straggler'' was in common use by the mid-1960s.

The much sparser cluster NGC 752\index{NGC 752} \cite{ro55} contains a single A0 star with a radial velocity and proper motion consistent with cluster membership. Arp \& Cuffey \cite{ar62b} compared the CMD for NGC 2158\index{NGC 2158} with those of NGC 7789 and NGC 752, the first group of ``intermediate-age clusters'' (ages from about one to a few billion years). They all have characteristic red giant clumps and contain BSSs. Other intermediate-age clusters with BSSs that have proper motions consistent with cluster membership include NGC 2477\index{NGC 2477} \cite{eg61} and NGC 6633\index{NGC 6633} \cite{hi58}.

\begin{figure}[h]
\sidecaption
\includegraphics[height=50mm]{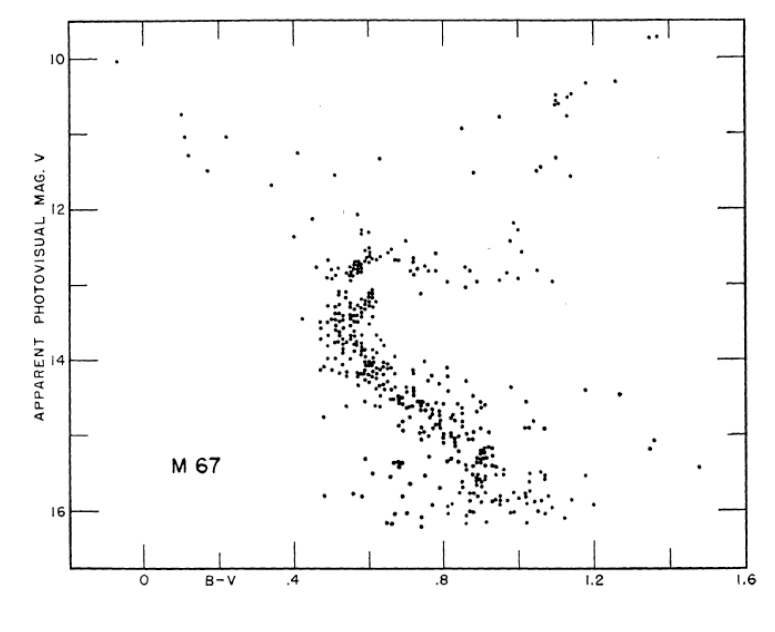}
\includegraphics[height=50mm]{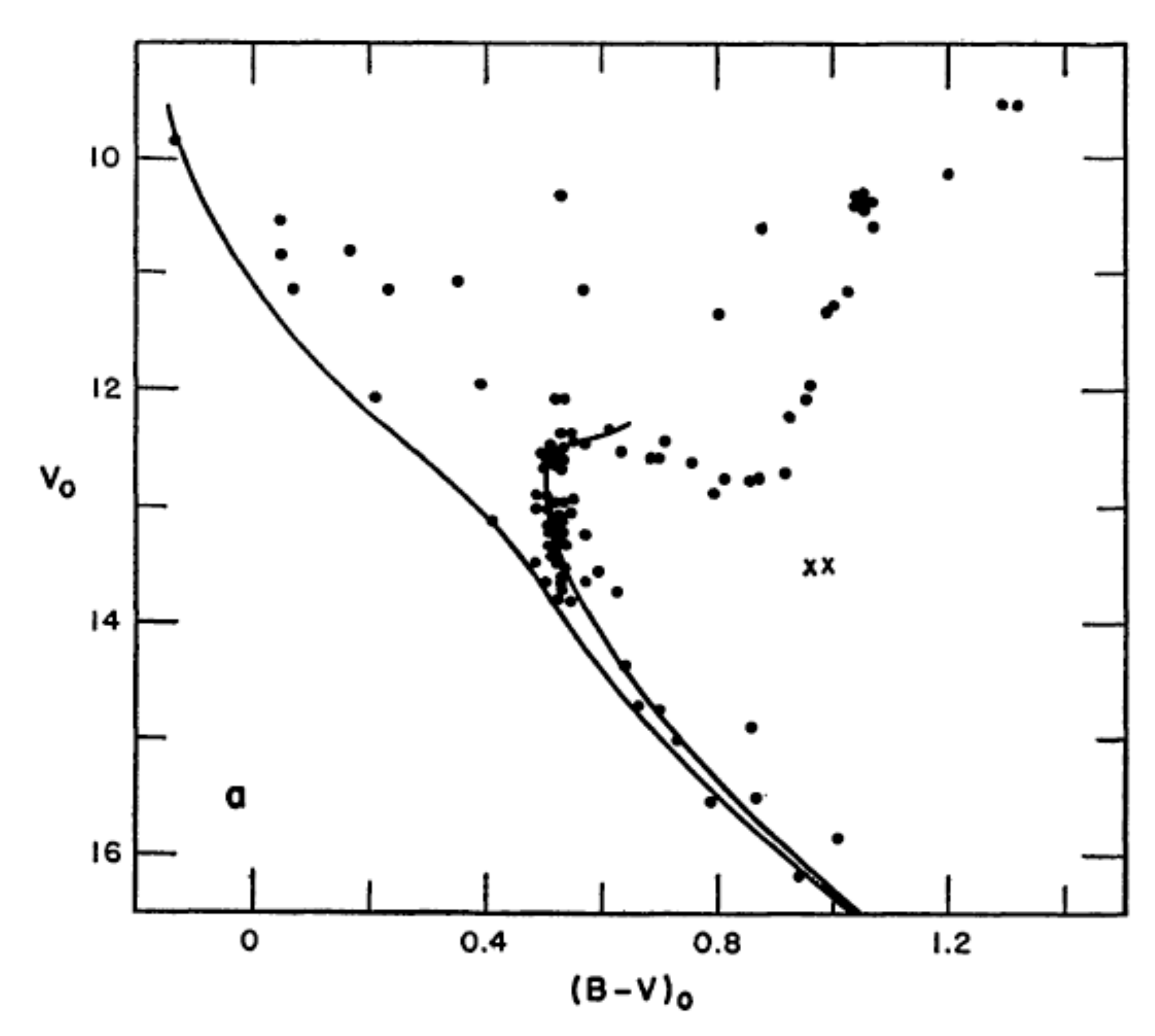}
\caption{CMDs of the old open cluster M67 by Johnson \& Sandage \cite{jo55} (left) and later (right) by Eggen \& Sandage \cite{eg64}, with improved photometry and some non-members omitted. Reprinted with permission from the AAS.}
\label{canfig3}       
\end{figure}

The oldest open clusters, such as M67\index{M67}, NGC 188\index{NGC 188} and NGC 6791\index{NGC 6791}, have ages between about five and ten billion years. They have CMDs qualitatively similar to those of globular clusters, with continuous subgiant and giant branches. M67 \cite{jo55} also contained blue stars, but in this case the situation was confused because the same stars could be interpreted as either anomalous main sequence stars or the analogues of globular cluster BHB stars (Fig.~\ref{canfig3}, left panel). Eggen \& Sandage \cite{eg64} produced a more accurate CMD (Fig.~\ref{canfig3}, right). In addition to improved photometry, they used proper motions\index{proper motion} \cite{mu65} to cull non-members from the sample. It is apparent that M67 contains a few stars that may be similar to the BSSs in M3, but it is difficult to distinguish between them and possible analogues of globular cluster BHB stars, as noted by Pesch \cite{pe67}, who checked that most of the potential BSSs had radial velocities consistent with cluster membership. Sargent \cite{sa68} also took spectra and found that the brightest BHB candidate in M67, star Fagerholm 81\index{Fagerholm 81} in the extreme top left corner of the plots in Fig.~\ref{canfig3}, had the gravity and hence high mass expected for a main sequence star at its location in the CMD, while the next three hottest stars had masses similar to stars on the upper main sequence, consistent with their being BHB stars. However, Bond \& Perry \cite{bo71b} obtained Str\"omgren photometry\index{Str\"omgren photometry} and disagreed with this latter conclusion, finding that all of these blue stars had relatively high masses and were genuine BSSs.

\begin{figure}[h]
\sidecaption
\includegraphics[width=119mm]{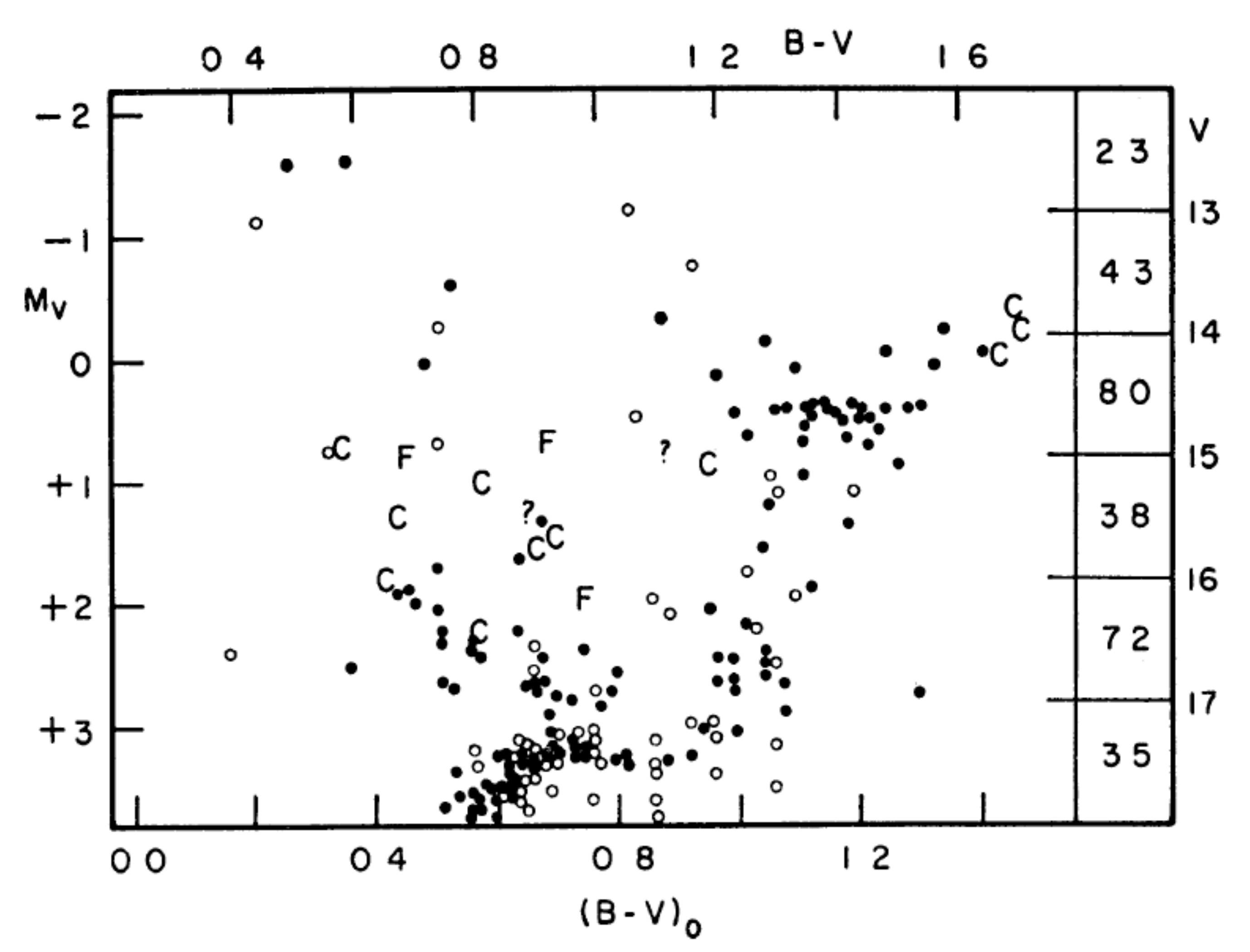}
\caption{The first CMD of the very old open cluster NGC 6791 \cite{ki65}. The letters ``C'' and ``F'' denote probable cluster members and field stars respectively, based on their radial velocities. Reprinted with permission from the AAS.}
\label{canfig4}       
\end{figure}

Sandage \cite{sa62} produced the first CMD of NGC 188\index{NGC 188}, one of the oldest open clusters. Its CMD is similar to that of M67 but with a fainter MSTO. This too had blue stars above the MSTO, with a larger magnitude gap between most of these and an ill-defined sparse clump of red giants so that there is less confusion with possible BHB stars. Sandage initially concluded that the blue stars were probably not cluster members because they did not seem to be concentrated towards the cluster centre. In a later paper with better photometry over a wider field, Eggen \& Sandage \cite{eg69} decided that at least some probably were BSS cluster members. NGC 188 is discussed in much more detail in Chap.3.

For my PhD work I studied a sample of old open clusters \cite{ca68}, using proper motions to remove field stars from their CMDs. This involved measuring matched pairs of photographic plates taken several decades apart. As often happens in astronomy, the program was determined by the available data. I was presented with a collection of photographs of star clusters from the Mount Wilson\index{Mount Wilson} 60-inch reflector, taken by Adriaan van Maanen circa 1920, together with modern repeats taken by Richard Woolley and Andrew Murray in 1961.

Eventually I had \emph{clean} CMDs for six clusters, all of which contained BSSs. These were M67\index{M67} \cite{mu65}, NGC 188\index{NGC 188} \cite{ca68}, NGC 6939\index{NGC 6939} \cite{ca69}, NGC 2420\index{NGC 2420} \cite{ca70a}, NGC 752\index{NGC 752} and NGC 7789\index{NGC 7789} (both unpublished). An analysis of the red giant clump stars was published \cite{ca70b} but the BSS analysis never was. The two main results were that virtually all of the BSSs lay within the Population I hydrogen-burning main sequence, between the zero-age main sequence (ZAMS) and the Sch\"onberg--Chandrasekhar limit, and almost all of them lay less than two magnitudes above the MSTO in each cluster. The BSSs were spread across the main sequence band, with no evidence for a concentration at the level of the horizontal branch. Thus all the data were consistent with them having formed by binary mass exchange\index{mass transfer}. Their distribution in magnitude and colour was consistent with simple models of the likely outcome of mass exchange, based on interpolation between a few published evolutionary tracks\index{evolutionary track} for single stars \cite{ib67} with two free parameters, the initial mass ratio\index{mass ratio} of the binary and the fraction of mass retained by the system.

The cluster NGC 6791\index{NGC 6791} is an interesting case, straddling the boundary between open and globular clusters. Kinman \cite{ki65} showed that it has a CMD similar to those of M67 and NGC 188 (Fig.~\ref{canfig4}). It is one of the oldest known open clusters \cite{me93}. However, its ``blue straggler'' sequence is more yellow than blue, being shifted significantly redward from the main sequence. The reality of this shift has been confirmed in a very similar CMD using modern CCD photometry \cite{mo94} and with cluster membership confirmed by proper motions \cite{cu93}. 

Very recently, NGC 6791 has received a great deal of attention, because it is the oldest known metal-rich cluster, with age $\sim$ 8 Gyr and [Fe/H] $\sim$ +0.4, and it lies in the region of sky covered by the NASA's \emph{Kepler} satellite. As such it was the subject of 25 published papers in 2011--2012. It contains BHB stars as well as BSSs, but perhaps most importantly the \emph{Kepler} data provide information on the core masses and mass loss from asteroseismology of individual stars \cite{mi12,co12}. 

\subsection{Younger Open Clusters and BSSs in the field}\label{cansec23}
It was realised early on that there were BSSs in younger clusters, but this branch has been more focused on the detailed study of individual relatively bright stars than on their CMDs. Few young clusters are populous enough to have many BSSs, and the ``classical'' definition based on the CMD only works when the spread in star formation times is small compared with the overall age of a cluster. 

Eggen \& Iben \cite{eg88,eg89a} looked for BSSs within ``moving groups'' of stars. This concept fell out of favour for some time, but is rapidly reappearing in the context of galaxy formation and the accretion of companion galaxies, and within the framework of ``Galactic Archaeology'' \cite{fr02}. A major compilation of BSSs in all open clusters has been put together by Ahumada \& Lapasset \cite{ah07}.

Blue stragglers must exist among the field stars in the Galaxy but finding them in the metal rich populations is difficult unless they have some characteristic other than being too young, for example as Algol-type systems (Sect.~\ref{cansec31}). However, they can be identified as hot metal-poor stars, as demonstrated for by Bond \& McConnell \cite{bo71a} and by George Preston (\cite{pr94}; see also Chap.4).

\section{Early Ideas on the Origin of Blue Stragglers}\label{cansec3}
To summarise, the essential properties of ``classical'' blue stragglers were that they are members of star clusters that lie on the main sequence\index{main sequence} but above the age-dependent turn-off defined by the majority of the stars in a cluster. Hence they seemed to be either younger or rejuvenated stars, or the consequence of some unusual process. 

The data described in Sect.~\ref{cansec2} illustrate some fundamental problems in pinning down BSSs. Globular clusters are very populous and can give large samples of rare types of star in well-defined locations in the CMD, and cluster membership can often be inferred only statistically. However, their BSSs are relatively faint, which may no longer be such a serious problem for photometry but still puts a limit on the amount and precision of high dispersion spectroscopic data. It is also difficult to obtain comparable data for complete samples of stars in the dense central cores of globular clusters and in the sparse outer regions. By contrast, open clusters are generally closer but much less populous, which means that better data can be obtained but often for only very small samples of BSSs. Open clusters also suffer more from interstellar dust and reddening, and from contamination by field stars so that accurate proper motions and radial velocities are needed to confirm cluster membership.

\subsection{Blue Stragglers and Algol-Type Eclipsing Binaries}\label{cansec31}

Although most BSSs were almost by definition members of clusters, it was quickly realised that they might be related to the Algol-type binary stars. These are short period eclipsing binaries in which the brighter and more massive star is located near the ZAMS\index{ZAMS}, while the fainter companion looks more highly evolved and hence older. However, they are too common to be the result of chance captures and must be coeval: the explanation is mass exchange\index{mass transfer}. 

Many astronomers of my generation were familiar with this concept thanks to a popular astronomy text by Fred Hoyle \cite{ho55}. It was presented in more detail by Crawford 
\cite{cr55} and by Kopal \cite{ko56}, who was puzzled by the preponderance of secondaries filling their Roche lobes\index{Roche lobe} in Algol\index{Algol system} systems. Morton \cite{mo60} made the first modern stellar evolution calculations and demonstrated that the mass transfer process would normally be unstable and hence short-lived. 

The possible link between BSSs and Algols was discussed at one of the first IAU Colloquia, on \emph{Star Clusters and Stellar Evolution}\footnote{A virtually verbatim transcript of the recorded proceedings of this meeting, held at the Royal Greenwich Observatory (RGO) while it was located at Herstmonceux Castle in Sussex, was published in 1964. The meeting is usually said to have been in 1963 but internal evidence points to 1962, a date recently confirmed by Bob Dickens (private communication, 2012). Although this and other Royal Observatory Bulletins are listed in the ADS, the text seems not to be available.} in 1962 \cite{eg64}.

The work of McCrea \cite{cannmc64} is probably the most cited reference for Algol-type mass exchange as the origin of BSSs; Smak \cite{sm66} developed similar ideas. Later, Algol variables were found among cluster BSSs. For example, Niss, Jorgensen \& Laustsen \cite{ni78} discovered the probable eclipsing variable NJL 5\index{NJL 5} in Omega Centauri\index{$\omega$ Centauri} and Margon \& Cannon \cite{ma89} showed that it was a radial velocity member of the cluster. However, repeated attempts to determine the velocity amplitude accurately or to detect the spectrum of the secondary component were unsuccessful. It was a bit too faint for the 4-m telescopes and spectrographs available at that time. 

\subsection{Other Possible Explanations for BSSs}\label{cansec32}
While binary mass exchange was the early favourite explanation for the formation of BSSs and remains one of the most popular today -- now strongly supported by a lot of evidence -- some stragglers do not fit that simple picture and many variants have been proposed. Fagerholm 81\index{Fagerholm 81}, the brightest BSS in M67\index{M67} (Fig. ~\ref{canfig3}), is almost three magnitudes above the MSTO and is too bright to have been formed from two main sequence stars. Presumably this required multiple mass transfers or collisions involving binary stars. Direct collisions between single stars are probably too rare to explain most BSSs, but Leonard \cite{cannle89} argued that binary star collisions should be much more efficient in low-density environments. Ferraro et al. \cite{cannfe93} provided convincing observational evidence for two distinct populations of BSSs in M3\index{M3}, one centrally concentrated and the other much more extended, which may well indicate two different formation mechanisms.

By comparison, several possible mechanisms that involved single stars were either discredited or believed to be much rarer: these included delayed or late star formation, stars whose main sequence lifetime had been extended by large scale mixing \cite{wh79}, highly evolved stars that happened to land close to the standard main sequence, tidally captured Galactic field stars, or Bondi-Hoyle gas accretion from the ISM. Most of the reviews cited in Sect.~\ref{cansec1} list the various hypotheses and discuss their relative merits.

\section{Expanding the Definition of BSSs, 1970--90}
\label{cansec4}
While observations of the relatively bright blue stragglers in open clusters dominated the literature up to about 1970, the emphasis then shifted towards globular clusters\index{globular cluster}. This was probably a consequence of the advent of a new generation of 4m-class telescopes, and then of CCDs from about 1980 onwards, which hugely increased the power of those telescopes. Coincidentally, the small area and slow readout of early CCDs made them difficult to use for accurate photometry of extended open clusters but they were well-matched to compact dense globular clusters. CCDs also opened up the field of spectroscopy\index{spectroscopy} of globular cluster stars, with a further large boost in telescope power when multi-fiber systems began to become available by 1990.

There were already fundamental flaws in the original definition of BSSs, since it only applied directly to stars in clusters and was not based on physically meaningful parameters. The term ``red giant'' works well because such stars are intrinsically redder (cooler) and bigger (larger radius) than most other stars. The ``blueness'' of blue stragglers is only relative to the MSTO, and many of the mechanisms devised to explain them can (and probably do) produce stars well below the turn-off that do not \emph{straggle}.

In addition, new discoveries gradually led to a complete revision of the importance of binary stars in star clusters, particularly in the cores of globular clusters. In the mid-1970s came the discovery of low-mass X-ray binaries\index{X-ray binary} and cataclysmic variables\index{cataclysmic variable} in many globular clusters \cite{po03}. A decade later it was found that millisecond pulsars\index{pulsar}, also believed to originate in binary systems, were exceptionally common in some clusters, most notably 47 Tucanae\index{47 Tucanae} \cite{ma01}. It became apparent that binary stars must be both created and destroyed in cluster cores. It also became clear that these mechanisms were of fundamental dynamical importance for understanding the evolution of clusters and the process of core collapse that had long puzzled theorists \cite{ba95}.

\section{Conclusions}\label{cansec5}
All these different perspectives on binary stars and BSSs mean that today probably everyone has a different mental picture of what is meant by the term ``blue straggler''.
It is to be hoped that one outcome of this book will be some guidelines on distinguishing between different types and specifying whether ``blue straggler'' is being used to define some particular stars or a formation process.

In parallel with the rapid evolution of ideas on BSSs, there has been a radical revision of ideas on the importance of the cluster environment for understanding the complex chemical abundance patterns seen within and between different clusters. The old paradigm that clusters were simple stellar systems that provided ideal tests for stellar evolution theory is no longer tenable. Although most clusters do seem to be made from material that is homogeneous in the relative abundance of iron and other heavy elements that dominate the opacity in the surface layers, there are large variations in the CNO group of elements and a few others such as Na. Moreover, very accurate faint photometry, mainly from HST, has shown that there are multiple stellar populations within some clusters. The exceptional cluster Omega Centauri (NGC 5139)\index{$\omega$ Centauri}\index{NGC 5139} is now regarded as perhaps the remnant core or bulge of a dwarf galaxy accreted by our Galaxy. It also appears that the abundances variations observed within many clusters are not representative of stars in the general field \cite{kr94}. Our standard theories of stellar evolution have been \emph{fine tuned} to fit globular clusters and may not be the whole story.

In any event, it will be difficult to synthesise high redshift galaxies without a complete prescriptive theory of star formation and evolution: BSSs are just one problem among several (the initial mass function; mass loss and horizontal branch structure; asymptotic giant branch evolution, supernovae and the return of processed material to the interstellar medium). Xin et al. \cite{xi11} have shown that BSSs can be numerous enough to contribute $\sim$10 percent to the integrated light of star clusters at short wavelengths (see also Chap.11 in this volume).

\backmatter
\printindex


\end{document}